\documentstyle[12pt]{article}
\begin{document}

\def\PRL{Phys.~Rev.~Lett.~}
\def\PR{Phys.~Rev.~}
\def\PRD{Phys.~Rev.~D}
\def\NP{Nucl.~Phys.~}
\def\NPB{Nucl.~Phys.~B}
\def\PL{Phys.~Lett.~}
\def\PLB{Phys.~Lett.~B}
\def\NC{Nouvo~Cim.~}

\def\Slash{\hskip -.6em/}
\def\ttbar{t\bar{t}}
\def\bbbar{b\bar{b}}
\def\btbar{b\bar{t}}
\def\tbbar{t\bar{b}}
\def\ImPi{{\rm Im}\Pi}
\def\Imlambda{{\rm Im}\lambda}
\def\ImDelta{{\rm Im}\Delta}
\def\Imf{{\rm Im}f}
\def\MSbar{\overline{\rm MS}}

\def\beq{\begin{equation}}
\def\eeq{\end{equation}}
\def\beqa{\begin{eqnarray}}
\def\eeqa{\end{eqnarray}}

\def\cf{{\it cf.}}
\def\ie{{\it i.e.}}

\begin{flushright}
FERMILAB-PUB-94/037-T \\
EFI 94-07 \\
hep-ph/9403294 \\
March 1994 \\
\end{flushright}

\bigskip
\medskip

\begin{center}

\large

{\bf Comment on the Dispersion Relations used to Calculate $\Delta\rho$.} \\

\bigskip
\medskip

\normalsize

\bigskip
{\it Tatsu Takeuchi \\
\medskip
Fermi National Accelerator Laboratory \\
P.O. Box 500, Batavia, IL 60510 } \\
\bigskip

and

\bigskip
{\it Aaron K. Grant and Mihir P. Worah} \\
\medskip
{\it Enrico Fermi Institute and Department of Physics \\
University of Chicago \\
5640 S. Ellis Avenue, Chicago, IL 60637} \\
\bigskip

\bigskip

%\newpage

{\bf ABSTRACT} \\
\end{center}

\begin{quote}
We use the operator product expansion (OPE) to show that
non-perturbative QCD
corrections to $\Delta\rho$ can be calculated using unsubtracted
dispersion relations for either the transverse or the longitudinal
vacuum polarization functions.
Recent calculations of the non-perturbative contribution to
$\Delta\rho$ based on a non-relativistic calculation of
corrections to the $\ttbar$ threshold are inconsistent with this result.

\end{quote}

\newpage

\centerline{\bf I.  INTRODUCTION}
\bigskip

An accurate calculation of the contribution of the $t$--$b$ doublet
to the $\rho$ parameter is important in two respects.
The first is that it will let us translate the tight constraint placed on
the $\rho$ parameter by LEP measurements to limits
on the $t$ quark mass, and the second is that
in the event that the $t$ quark mass is measured directly at the
TEVATRON, it will help us constrain the contribution of new physics
to $\Delta\rho$.

The contribution of the $t$--$b$ doublet to the $\rho$ parameter
has been calculated to $O(\alpha\alpha_s)$, and the result is given by
\beq
\Delta\rho =  \frac{3\alpha m_t^2}{16\pi s^2 c^2 m_Z^2}
              \left[ 1 - \frac{\alpha_s}{\pi}
                         \left( \frac{2\pi^2+ 6}{9}
                         \right)
              \right]
\label{TwoLoop}
\eeq
in the limit $m_b \rightarrow 0$ \cite{DV,Kniehl}.

Recently, Kniehl and Sirlin  estimated the
size of the higher order QCD corrections to Eq.~(\ref{TwoLoop})
using a dispersion relation for $\Delta\rho$ \cite{KS}.
Their approach has been to assume that the effect
of non--perturbative QCD corrections is dominated by the
change in the shape of the $\ttbar$ threshold.   This change
is then calculated using the leading non--relativistic approximation,
and substituted into the dispersion relation for $\Delta\rho$.
This work has lead to a certain amount of controversy since different
authors found the effect to be different in magnitude,
ranging from 10\% to 80\% of the $O(\alpha\alpha_s)$ correction,
and sometimes even different in sign \cite{Yndurain,Halzen}.

The difference in sign comes from the fact the there are two possible
dispersion relations for $\Delta\rho$.
In order to explain what they are, we must introduce some notation.
Following Ref.~\cite{KS}, we define
\beqa
\Pi^{V,A}_{\mu\nu}(q,m_1,m_2)
& = & -i \int d^4x e^{iq\cdot x}
\langle 0 | T^* \left[ J_\mu^{V,A}(x) J_\nu^{V,A\dagger}(0)
                \right]
| 0 \rangle                                            \nonumber \\
& = & g_{\mu\nu} \Pi^{V,A}(s,m_1,m_2)
    + q_\mu q_\nu \lambda^{V,A}(s,m_1,m_2)
      \phantom{\frac{1}{2}}                            \nonumber \\
& = & \left( g_{\mu\nu} - \frac{q^\mu q^\nu}{q^2}
      \right) \Pi^{V,A}(s,m_1,m_2)
    + \left( \frac{q^\mu q^\nu}{q^2}
      \right) \Delta^{V,A}(s,m_1,m_2)  \nonumber \\
\label{Defs1}
\eeqa
where $s = q^2$, and $J_\mu^{V,A}(x)$ represents
the vector and axial vector currents
constructed from the fermion fields, respectively.
Note that
\beqa
\Delta^{V,A}(s) = \Pi^{V,A}(s) + s\lambda^{V,A}(s)
\label{Defdelm}
\eeqa
so that
\beq
\Delta^{V,A}(0) = \Pi^{V,A}(0),
\eeq
unless $\lambda^{V,A}(s)$ has a pole at $s=0$.
We further introduce the notation
\beqa
\Pi^{V,A}_\pm(s)
& = & \Pi^{V,A}(s,m_1,m_2),     \phantom{\frac{1}{2}} \nonumber \\
\lambda^{V,A}_\pm(s)
& = & \lambda^{V,A}(s,m_1,m_2), \phantom{\frac{1}{2}} \nonumber \\
\Delta^{V,A}_\pm(s)
& = & \Delta^{V,A}(s,m_1,m_2),  \phantom{\frac{1}{2}} \nonumber \\
\Pi^{V,A}_0(s)
& = & \frac{1}{2} \left[ \Pi^{V,A}(s,m_1,m_1) + \Pi^{V,A}(s,m_2,m_2)
                  \right],       \nonumber \\
\lambda^{V,A}_0(s)
& = & \frac{1}{2} \left[ \lambda^{V,A}(s,m_1,m_1) + \lambda^{V,A}(s,m_2,m_2)
                  \right],       \nonumber \\
\Delta^{V,A}_0(s)
& = & \frac{1}{2} \left[ \Delta^{V,A}(s,m_1,m_1) + \Delta^{V,A}(s,m_2,m_2)
                  \right].
\label{Defs2}
\eeqa
The conservation of the neutral vector currents implies
the Ward Identities:
\beq
\Pi^V_0(s) = -s\lambda_0(s), \qquad\qquad \Delta^V_0(s) \equiv 0.
\label{WI}
\eeq
These definitions let us write
the contribution of an $SU(2)$ fermion doublet,
with masses $m_1$ and $m_2$,
to the vacuum polarizations of the
$W$ and the $Z$ at zero momentum transfer $s=0$  as
\beqa
\Pi_{WW}(0) & = & \frac{g^2}{8}
                \left[ \Pi^V_\pm(0) + \Pi^A_\pm(0)
                \right]
              =   \frac{g^2}{8}
                \left[ \Delta^V_\pm(0) + \Delta^A_\pm(0)
                \right],    \nonumber \\
\Pi_{ZZ}(0) & = & \frac{g^2+g'^2}{8}
                \left[ \Pi^V_0(0) + \Pi^A_0(0)
                \right]
              =   \frac{g^2+g'^2}{8} \Delta^A_0(0).
\label{WWandZZ}
\eeqa
Note that $\Pi^V_0(0)$ is actually zero from current conservation
(\cf~Eq.~(\ref{WI}))
but we will keep it in our expressions for later convenience.
Inserting Eq.~(\ref{WWandZZ}) into the definition of the $\rho$ parameter,
we find
\beqa
\Delta\rho
& = & \frac{ \Pi_{WW}(0) }{ M_W^2 } - \frac{ \Pi_{ZZ}(0) }{ M_Z^2 }
\nonumber  \\
& = & \frac{ G_F }{ \sqrt{2} }
      \left\{ \left[ \Pi^V_\pm(0) + \Pi^A_\pm(0)
              \right]
            - \left[ \Pi^V_0(0)   + \Pi^A_0(0)
              \right]
      \right\} \nonumber \\
& = & \frac{ G_F }{ \sqrt{2} }
      \left\{ \left[ \Delta^V_\pm(0) + \Delta^A_\pm(0)
              \right] - \Delta^A_0(0)
      \right\}
\label{Delrho}
\eeqa
Applying the `unsubtracted' dispersion relation:
\beq
f(s) = \frac{1}{\pi}\int^\infty ds' \frac{ \Imf(s') }{ s'-s+i\epsilon }
\label{fDR}
\eeq
to the expression for $\Delta\rho$ using the $\Pi(s)$'s gives us
\beq
\Delta\rho
= \frac{ G_F }{ \sqrt{2} }
  \frac{1}{\pi} \int^\infty \frac{ds}{s}
  \left[ \left\{ \ImPi^V_\pm(s) + \ImPi^A_\pm(s)
          \right\}
        - \left\{ \ImPi^V_0(s) + \ImPi^A_0(s)
          \right\}
  \right].
\label{NaiveDR}
\eeq
This dispersion relation is equivalent to that
introduced in Ref.~\cite{Chang},
and used in Ref.~\cite{Kniehl} to calculate the $O(\alpha\alpha_s)$
correction.
%and corresponds to
%using unsubtracted dispersion relations for the transverse parts of
%the vacuum polarizations that appear in the expression for $\Delta\rho$.

If we write down an unsubtracted dispersion relation for the expression of
$\Delta\rho$ using the $\Delta(s)$'s, we get
\beqa
\Delta\rho
= \frac{ G_F }{ \sqrt{2} }
  \frac{1}{\pi} \int^{\infty} \frac{ds}{s}
  \left[ \left\{ \ImDelta^V_\pm(s) + \ImDelta^A_\pm(s)
         \right\} -\ImDelta^A_0(s)
  \right].
\label{KSDR}
\eeqa
This is the relation that was introduced in Ref.~\cite{KS}.
Eqs.~(\ref{NaiveDR}) and (\ref{KSDR}) are the two dispersion
relations for $\Delta\rho$ that have appeared in the literature.  Our point
here is to illustrate that, although Eqs.~(\ref{NaiveDR},~\ref{KSDR}) should
yield the same estimate for $\Delta\rho$, under certain commonly used
approximations they do not.

In Sec. II  we clarify the conditions that
the $\Pi(s)$'s and $\Delta(s)$'s must satisfy for
Eqs.~(\ref{NaiveDR}) and (\ref{KSDR}) to be true. These conditions are
indeed satisfied to order $\alpha\alpha_s$ in perturbation theory. In
Sec. III,
we argue that
the operator product expansion (OPE) of the current--current
correlators suggests that these conditions are
satisfied for non--perturbative QCD corrections so that
both Eqs.~(\ref{NaiveDR}) and (\ref{KSDR}) are valid. In Sec. IV we
show that the two dispersion relations Eqs.~(\ref{NaiveDR}) and
(\ref{KSDR}) give different answers when used in a threshold
approximation to calculate
non-perturbative effects. We argue that the
 disagreement between the two dispersion relations can
be understood as a result of neglecting the non--threshold contribution
of the $\ImPi(s)$'s and $\ImDelta(s)$'s to $\Delta\rho$.  Furthermore,
the magnitude of the discrepancy between Eqs.~(\ref{NaiveDR},~\ref{KSDR})
can be used
as a measure of the accuracy of the calculation. We conclude in Sec. V.

\bigskip
\centerline{\bf II. THE DISPERSION RELATIONS FOR $\Delta\rho$}
\bigskip

Let us begin this section by recalling that
the analyticity of a vacuum polarization function $f(s)$ and
Cauchy's theorem tell us that
\beq
f(s) = \frac{1}{\pi}\int^{\Lambda^2} ds'
       \frac{ \Imf(s') }{ s'-s+i\epsilon }
     + \frac{1}{2\pi i}\oint_{|s|=\Lambda^2} ds'
       \frac{  f(s')   }{ s'-s }.
\label{fCauchy}
\eeq
Hence for Eq.~(\ref{fDR}) to be true, $f(s)$ must satisfy
\beq
\lim_{\Lambda^2\rightarrow\infty}
\oint_{|s|=\Lambda^2} \frac{ f(s') }{ s'-s } = 0.
\label{fCondition}
\eeq
Applying Eq.~(\ref{fCauchy}) to the $\Pi(0)$'s in Eq.~(\ref{Delrho}),
we find
\beqa
\Pi^V_\pm(0)
& = & \frac{1}{\pi}\int^{\Lambda^2} \frac{ds}{s} \ImPi^V_\pm(s)
    + \frac{1}{2\pi i} \oint_{|s|=\Lambda^2} \frac{ds}{s} \Pi^V_\pm(s),
      \nonumber \\
\Pi^A_\pm(0)
& = & \frac{1}{\pi}\int^{\Lambda^2} \frac{ds}{s} \ImPi^A_\pm(s)
    + \frac{1}{2\pi i} \oint_{|s|=\Lambda^2} \frac{ds}{s} \Pi^A_\pm(s),
      \nonumber \\
\Pi^V_0(0)
& = & \frac{1}{\pi}\int^{\Lambda^2} \frac{ds}{s} \ImPi^V_0(s)
    + \frac{1}{2\pi i} \oint_{|s|=\Lambda^2} \frac{ds}{s} \Pi^V_0(s),
      \nonumber \\
\Pi^A_0(0)
& = & \frac{1}{\pi}\int^{\Lambda^2} \frac{ds}{s} \ImPi^A_0(s).
    + \frac{1}{2\pi i} \oint_{|s|=\Lambda^2} \frac{ds}{s} \Pi^A_0(s).
\label{PiCauchy}
\eeqa
We regard the $\Pi$'s in these relations to be {\it regularized}
\footnote{With $\MSbar$ or some other regularization scheme
which respects the symmetry of the theory so that the Ward Identities
are satisfied.}
and thus finite quantities so that both the left and right hand sides
of these equations are well defined.
Note that the Ward Identity, Eq.~(\ref{WI}), ensures that
\beqa
\Pi^V_0(0)
& = & \frac{1}{\pi}\int^{\Lambda^2} \frac{ds}{s} \ImPi^V_0(s)
    + \frac{1}{2\pi i} \oint_{|s|=\Lambda^2} \frac{ds}{s} \Pi^V_0(s),
      \nonumber \\
& = & - \left[
        \frac{1}{\pi}\int^{\Lambda^2} ds \Imlambda^V_0(s)
      + \frac{1}{2\pi i} \oint_{|s|=\Lambda^2} ds \lambda^V_0(s),
        \right]
      \nonumber \\
& = & - \frac{1}{2\pi i}\oint ds \lambda^V_0(s) \nonumber    \\
& = & 0,
\label{WICauchy}
\eeqa
as required.  This result follows simply from Cauchy's theorem and the
(assumed) analyticity of $\lambda(s)$.
If we define
\beqa
\Delta\rho_T(\Lambda^2)
& \equiv &
\frac{G_F}{\sqrt{2}}\frac{1}{\pi}
\int^{\Lambda^2} \frac{ds}{s}
\left[ \left\{ \ImPi^V_\pm(s) + \ImPi^A_\pm(s)
       \right\}
     - \left\{ \ImPi^V_0(s)   + \ImPi^A_0(s)
       \right\}
\right],                \nonumber \\
\Delta R_T(\Lambda^2)
& \equiv &
\frac{G_F}{\sqrt{2}}\frac{1}{2\pi i}
\oint_{|s|=\Lambda^2} \frac{ds}{s}
\left[ \left\{ \Pi^V_\pm(s) + \Pi^A_\pm(s)
       \right\}
     - \left\{ \Pi^V_0(s)   + \Pi^A_0(s)
       \right\}
\right],                \nonumber \\
\label{DefsT}
\eeqa
the substitution of Eq.~(\ref{PiCauchy}) into Eq.~(\ref{Delrho})
gives
\beq
\Delta\rho = \Delta\rho_T(\Lambda^2) + \Delta R_T(\Lambda^2).
\label{DelrhoT}
\eeq
(The subscript ``$T$'' stands for ``transverse''.)
Therefore, for the dispersion relation Eq.~(\ref{NaiveDR}) to be
valid, we must have
\beq
\lim_{\Lambda^2\rightarrow\infty}
\Delta R_T(\Lambda^2) = 0.
\label{ConditionT}
\eeq
This requires the linear combination of the $\Pi(s)$'s in the
integrand of $\Delta R_T(\Lambda^2)$ to vanish as $|s|\rightarrow\infty$.

Similarly, applying Eq.~(\ref{fCauchy}) to the $\Delta(s)$'s gives us
\beqa
\Delta^V_\pm(0)
& = & \frac{1}{\pi}\int^{\Lambda^2} \frac{ds}{s} \ImDelta^V_\pm(s)
    + \frac{1}{2\pi i} \oint_{|s|=\Lambda^2} \frac{ds}{s} \Delta^V_\pm(s),
      \nonumber \\
\Delta^A_\pm(0)
& = & \frac{1}{\pi}\int^{\Lambda^2} \frac{ds}{s} \ImDelta^A_\pm(s)
    + \frac{1}{2\pi i} \oint_{|s|=\Lambda^2} \frac{ds}{s} \Delta^A_\pm(s),
      \nonumber \\
\Delta^A_0(0)
& = & \frac{1}{\pi}\int^{\Lambda^2} \frac{ds}{s} \ImDelta^A_0(s).
    + \frac{1}{2\pi i} \oint_{|s|=\Lambda^2} \frac{ds}{s} \Delta^A_0(s).
\label{DeltaCauchy}
\eeqa
Again, we consider the $\Delta(s)$'s to be regularized and finite
quantities. If we define
\beqa
\Delta\rho_L(\Lambda^2)
& \equiv &
\frac{G_F}{\sqrt{2}}\frac{1}{\pi}
\int^{\Lambda^2} \frac{ds}{s}
\left[ \left\{ \ImDelta^V_\pm(s) + \ImDelta^A_\pm(s)
       \right\}
     - \ImDelta^A_0(s)
\right],                \nonumber \\
\Delta R_L(\Lambda^2)
& \equiv &
\frac{G_F}{\sqrt{2}}\frac{1}{2\pi i}
\oint_{|s|=\Lambda^2} \frac{ds}{s}
\left[ \left\{ \Delta^V_\pm(s) + \Delta^A_\pm(s)
       \right\}
     - \Delta^A_0(s)
\right],                \nonumber \\
\label{DefsL}
\eeqa
the substitution of Eq.~(\ref{DeltaCauchy}) into Eq.~(\ref{Delrho})
gives us
\beq
\Delta\rho = \Delta\rho_L(\Lambda^2) + \Delta R_L(\Lambda^2).
\label{DelrhoL}
\eeq
(The subscript ``$L$'' stands for ``longitudinal''.)
Therefore, for the dispersion relation Eq.~(\ref{KSDR}) to be
valid, we must have
\beq
\lim_{\Lambda^2\rightarrow\infty}
\Delta R_L(\Lambda^2) = 0,
\label{ConditionL}
\eeq
which requires the linear combination of the $\Delta(s)$'s in the
integrand of $\Delta R_L(\Lambda^2)$ to vanish as $|s|\rightarrow\infty$.

The actual asymptotic forms of the
$\Pi(s)$'s and $\Delta(s)$'s up to order $\alpha\alpha_s$
 can by found in Ref.~\cite{DG}
and one can explicitly check that
Eqs.~(\ref{ConditionT}) and (\ref{ConditionL}) hold.
Therefore, to order $\alpha\alpha_s$ in perturbation theory,
\beq
  \Delta\rho
= \Delta\rho_T(\infty)
= \Delta\rho_L(\infty),
\eeq

\bigskip
\centerline{\bf III. THE OPERATOR PRODUCT EXPANSION}
\bigskip

In the previous section, we have seen that
\beq
\Delta R_T(\infty) = \Delta R_L(\infty) = 0
\label{ConditionX}
\eeq
implies
\beq
\Delta\rho = \Delta\rho_T(\infty) = \Delta\rho_L(\infty),
\label{XXX}
\eeq
and that Eq.~(\ref{ConditionX}) holds
to order $O(\alpha\alpha_s)$ in perturbation theory.
Whether these equations continue to be valid
to higher orders in perturbation theory, and non-perturbatively
is a difficult problem in general.
However, since we are only interested in QCD corrections,
the non--perturbative asymptotic behavior of the vacuum polarization functions
$\Pi(s)$ and
$\Delta(s)$ as $|s|\rightarrow\infty$
can still be extracted from their operator product expansions.

The OPE's for $\Pi_{\pm,0}(s)$ and $\Delta_{\pm,0}(s)$ can be found
in the appendix of Ref.~\cite{Braaten}.
They are:
\beqa
\Delta^V_\pm(-Q^2)
& = & \hat{C}_{\Delta 1}(Q)
      \left[ \hat{m}_1(Q) - \hat{m}_2(Q)
      \right]^2
    + \hat{C}_{\Delta 2}(\mu)
      \left[ \hat{m}_1(\mu) - \hat{m}_2(\mu)
      \right]^2
    + O\left( \frac{1}{Q^2}
       \right),
\nonumber \\
\Delta^A_\pm(-Q^2)
& = & \hat{C}_{\Delta 1}(Q)
      \left[ \hat{m}_1(Q) + \hat{m}_2(Q)
      \right]^2
    + \hat{C}_{\Delta 2}(\mu)
      \left[ \hat{m}_1(\mu) + \hat{m}_2(\mu)
      \right]^2
    + O\left( \frac{1}{Q^2}
       \right),
\nonumber \\
\Delta^A_0(-Q^2)
& = & \hat{C}_{\Delta 1}(Q)
      \left[ 2\hat{m}_1(Q)^2   + 2\hat{m}_2(Q)^2
      \right]
\nonumber \\
& & + \hat{C}_{\Delta 2}(\mu)
      \left[ 2\hat{m}_1(\mu)^2 + 2\hat{m}_2(\mu)^2
      \right]
    + O\left( \frac{1}{Q^2}
       \right),
\label{DeltaOPE}
\eeqa
and
\beq
\Pi^{V,A}_{\pm,0}(-Q^2) = Q^2\lambda^{V,A}_{\pm,0}(-Q^2) +
\Delta^{V,A}_{\pm,0}(-Q^2)
\label{PiOPE}
\eeq
where
\beqa
\lambda^V_\pm(-Q^2)
& = & \hat{C}_{\lambda1}(Q)
    + \hat{C}_{\lambda2}(Q)\frac{[\hat{m}_1(Q) + \hat{m}_2(Q)]^2}{Q^2}
      \nonumber \\
&   & \qquad\qquad
    + \hat{C}_{\lambda3}(Q)\frac{[\hat{m}_1(Q) - \hat{m}_2(Q)]^2}{Q^2}
    + O\left( \frac{1}{Q^4} \right),
      \nonumber \\
\lambda^A_\pm(-Q^2)
& = & \hat{C}_{\lambda1}(Q)
    + \hat{C}_{\lambda2}(Q)\frac{[\hat{m}_1(Q) - \hat{m}_2(Q)]^2}{Q^2}
      \nonumber \\
&   & \qquad\qquad
    + \hat{C}_{\lambda3}(Q)\frac{[\hat{m}_1(Q) + \hat{m}_2(Q)]^2}{Q^2}
    + O\left( \frac{1}{Q^4} \right),
      \nonumber \\
\lambda^V_0(-Q^2)
& = & \hat{C}_{\lambda1}(Q)
    + \hat{C}_{\lambda2}(Q)\frac{[2\hat{m}_1(Q)^2 + 2\hat{m}_2(Q)^2]}{Q^2}
    + O\left( \frac{1}{Q^4} \right),
      \nonumber \\
\lambda^A_0(-Q^2)
& = & \hat{C}_{\lambda1}(Q)
    + \hat{C}_{\lambda3}(Q)\frac{[2\hat{m}_1(Q)^2 + 2\hat{m}_2(Q)^2]}{Q^2}
    + O\left( \frac{1}{Q^4} \right).
      \nonumber \\
\label{lambdaOPE}
\eeqa
The short distance physics is embodied in the Wilson Coefficients
$\hat{C}_*(Q)$ which can be calculated perturbatively, and their
explicit forms for the first few orders in $\alpha_s(Q)$ can be found
in Ref.~\cite{Braaten}.  The long distance non--perturbative physics
is embodied in the running masses $\hat{m}_{1,2}(Q)$ and the vacuum expectation
values of the higher dimensional operators that appear at higher order in
the OPE.   The Wilson coefficients are independent of the
masses except through the running of $\alpha_s(Q)$.

That only these combinations of running masses appear at dimension 2
in the OPE can be understood as follows:
Consider the charged channel functions
$\Delta^{V,A}_\pm(s)$ and $\lambda^{V,A}_\pm(s)$.  Since they must be
symmetric under the interchange $\hat{m}_1 \leftrightarrow \hat{m}_2$,
they can only depend on $(\hat{m}_1 \pm \hat{m}_2)^2$.
Changing the sign of one of the masses will interchange the vector
and axial--vector cases so the coefficient of $(\hat{m}_1 \pm \hat{m}_2)^2$
in the vector channel is equal to the coefficient of
$(\hat{m}_1 \mp \hat{m}_2)^2$ in the axial--vector channel.
Since $\Delta^V_\pm(s)$ must vanish when $\hat{m}_1 = \hat{m}_2$, it can
only depend on $(\hat{m}_1 - \hat{m}_2)^2$,
which in turn means that $\Delta^A_\pm(s)$ can depend only on
$(\hat{m}_1 + \hat{m}_2)^2$.
The dependence of the neutral channel functions
$\Delta^{V,A}_0(s)$ and $\lambda^{V,A}_0(s)$
on the running masses follows trivially from that of the
charged channel functions.

Note that the Wilson Coefficients $\hat{C}_*(Q)$ and the
running masses $\hat{m}_{1,2}(Q)$ depend only logarithmically
on $Q$.
Therefore,
though these OPE's are derived in the deep Euclidean region $-s = Q^2 \gg 0$,
we can expect the dependence of the $\Delta(s)$'s  and $\lambda(s)$'s
on the powers of $s$ to be the same
all around the circle $|s|=\Lambda^2$.
We can see immediately that this implies
$\Delta R_L(\infty) = \Delta R_T(\infty) = 0$.
We can therefore conclude that
Eq.~(\ref{XXX}) is correct even when higher order and
non--perturbative QCD corrections are taken into account.

\bigskip
\centerline{\bf IV. CALCULATING $\ttbar$ THRESHOLD EFFECTS}
\bigskip

Let us now turn to the problem of using dispersion relations to
calculate the
non--perturbative QCD corrections to $\Delta\rho$.
In order to make use of the
dispersion relations $\Delta\rho = \Delta\rho_T(\infty)$
and/or $\Delta\rho = \Delta\rho_L(\infty)$,
we need to know the spectral functions $\ImPi(s)$ and/or
$\ImDelta(s)$ when non--perturbative corrections are taken into account.
Since it is impossible to calculate them exactly,
this means that we must make some assumptions and approximations
about their non--perturbative behavior.

Let us denote the difference between the non--perturbative and
perturbative vacuum polarization functions, by
\beqa
\delta\ImPi(s)     & \equiv & \ImPi_{NP}(s)     - \ImPi_{P}(s),
\nonumber \\
\delta\ImDelta(s)  & \equiv & \ImDelta_{NP}(s)  - \ImDelta_{P}(s),
\eeqa
In Ref.~\cite{KS}, it was assumed that the most important
effect of non--perturbative QCD corrections
on $\ImPi(s)$ and $\ImDelta(s)$
is to modify the shape of the $\ttbar$ threshold.
The threshold region is also the place where higher order corrections
in $\alpha_s$ can be resummed in the leading non--relativistic
approximation and calculated reliably in terms of a simple non-relativistic
Schr\"odinger Green's function, as in Ref.~\cite{SPeskin}.
Far from the threshold region, the $\ImPi(s)$'s  are
assumed to be well approximated by their $O(\alpha\alpha_s)$ perturbative
results.
Therefore, in this approximation, the functions $\delta\ImPi(s)$ and
$\delta\ImDelta(s)$
have their support only in the region
near the threshold.
Furthermore, in the leading non--relativistic approximation,
 only the $s$--wave states contribute so that
\beq
            \delta_{\ttbar}\ImPi^V_0(s)
   =     - s\delta_{\ttbar}\Imlambda^V_0(s)
   =     - s\delta_{\ttbar}\Imlambda^A_0(s)
   =     -  \delta_{\ttbar}\ImDelta^A_0(s),
\label{NR}
\eeq
while all the other $\delta_{\ttbar}\ImPi(s)$'s,
$\delta_{\ttbar}\ImDelta(s)$'s, and
$\delta_{\ttbar}\Imlambda(s)$'s are zero.
(We have added a subscript to $\delta$ to indicate that since we don't
expect large threshold corrections in the $\bbbar$ or $\tbbar$
channels, only
$\ttbar$ threshold effects are included. This would not be the case if
we were considering threshold effects for say a heavy fourth
generation of quarks, but including the other
 channels does not alter our conclusions).
Note that the first equality in Eq.~(\ref{NR}) comes from
the Ward Identity Eq.~(\ref{WI}),
the second equality comes from the spin independence of
QCD interactions in the non--relativistic limit,
and the third equality comes
from Eq.~(\ref{Defdelm}), and
the fact that $\delta_{\ttbar}\ImPi^A_0(s) = 0$
in this approximation.
We will not specify what these non--zero terms look like
in any detail since it is irrelevant to the following discussion.

Now, let us see what happens when we
substitute Eq.~(\ref{NR}) into the definitions of
$\Delta\rho_T(\infty)$ and $\Delta\rho_L(\infty)$.
We find:
\beq
\delta_{\ttbar}
\left[ \Delta\rho_T(\infty)
\right]
 = - \frac{G_F}{\sqrt{2}}
     \left[
     \frac{1}{\pi} \int^\infty \frac{ds}{s} \delta_{\ttbar}\ImPi^V_0(s)
     \right]
 \equiv X,
\label{CASE1}
\eeq
and
\beq
\delta_{\ttbar}
\left[ \Delta\rho_L(\infty)
\right]
 = - \frac{G_F}{\sqrt{2}}
     \left[ \frac{1}{\pi} \int^\infty \frac{ds}{s}
            \delta_{\ttbar}\ImDelta^A_0(s)
     \right]
 =   \frac{G_F}{\sqrt{2}}
     \left[
     \frac{1}{\pi} \int^\infty \frac{ds}{s}
     \delta_{\ttbar}\ImPi^V_0(s)
     \right]
 = -X,
\label{CASE2}
\eeq
which are of the same magnitude but opposite in sign.
This is the disagreement in sign that was mentioned in the
introduction.

There are two ways to interpret this result.
The first is that
 either one, or both of the dispersion
relations Eq.~(\ref{NaiveDR}) and (\ref{KSDR}) are wrong.
This means that Eqs.~(\ref{ConditionT}) and (\ref{ConditionL})
cannot be both correct. This is the approach adopted by the authors of
Ref.~\cite{KS} who, for a number of reasons,
prefer Eq.~(\ref{KSDR}) to Eq.~(\ref{NaiveDR}).
However, we do not think this possibility is very
likely as we don't see where the OPE argument of the previous section
could have failed.

A second and more plausible possibility is that using only the leading
non--relativistic limit to calculate non--perturbative contributions to
$\Delta\rho$ is simply not a good approximation
and that the disagreement between Eq.~(\ref{CASE1}) and (\ref{CASE2})
is a reflection of that fact.

Let us take a closer look at where this apparent inconsistency between
the two approaches comes from. The reason we get
the same magnitude but opposite signs in Eqs.~(\ref{CASE1}) and
(\ref{CASE2}) is because of Eq.~(\ref{NR}), which is only true
at leading order in the non--relativistic approximation
in a small region near the threshold.
If we could increase the range where we can
calculate non-perturbative effects, Eq.~(\ref{NR}) will not be true
over the entire integration interval
and we expect the difference between the two results to be reduced.

We illustrate this with an example:
Consider the non-relativistic limit of the vacuum polarizations
calculated to one loop in the $\ttbar$ channel,
\beqa
\frac{\ImPi_{0,\ttbar}^V(s)}{s} & = & -\frac{3 \beta}{16\pi} + O(\beta^3),
\nonumber \\
\frac{\ImPi_{0,\ttbar}^A(s)}{s} & = & O(\beta^3), \nonumber\\
\frac{\ImDelta_{0,\ttbar}^A(s)}{s} & = & \frac{3 \beta}{16\pi} + O(\beta^3),
\label{NRPert}
\eeqa
where $\beta = \sqrt{1-4m_t^2/s}$. So, if we decided to
calculate the perturbative contribution to $\Delta\rho$ using the
leading non-relativistic approximation
 (and including all the other channels), we would again get answers of
the same magnitude but opposite sign depending on whether we chose to
calculate using the $\Pi(s)$'s or the $\Delta(s)$'s. However, from the
full perturbative calculation, we know that both techniques should
give the same answer. For similar reasons, (although there are
important differences in the perturbative and non-perturbative cases),
we believe pushing the cutoff further and further
away from the threshold would cause the non perturbative
results obtained using either the
$\Pi(s)$'s or the $\Delta(s)$'s to converge towards each other,
though perhaps very slowly.
(The importance of the contribution from regions away from the
threshold has already been noted in Ref.~\cite{Halzen}.)
We are currently studying how much of this convergence can be achieved
by taking the non--relativistic approximation to higher orders in $\beta$
and thus expanding its region of applicability.

In any approximation, a good way to test its accuracy is to see how
well it reproduces a known result.   In the present case,
a good way to test how well our approximation gives the
correct value for $\Delta\rho_T(\infty)$ or $\Delta\rho_L(\infty)$ is
to see how well it reproduces the known value for
$\Delta\rho_T(\infty) - \Delta\rho_L(\infty)$, namely zero.
When seen in this context, Eqs.~(\ref{CASE1},~\ref{CASE2})
are an indication that
the $\ttbar$ threshold approximation fails, indeed we cannot even
determine the sign of the additional contribution to $\Delta\rho$.

Of course, the bright side of it is that now we may understand
the reason why the two dispersion relations Eq.~(\ref{NaiveDR})
and Eq.~(\ref{KSDR}) give seemingly contradicting results
for the $\ttbar$ threshold effects.   In fact, they are not contradicting
at all.   The difference between the two values is the error
that should be associated with the approximation.

\bigskip
\centerline{\bf V. CONCLUSION}
\bigskip

We have used the OPE to show that it is equally valid to calculate
$\Delta\rho$ using either the dispersion relations of
Ref.~\cite{Chang}, or those of Ref.~\cite{KS}. These correspond to
using unsubtracted dispersion relations for either the transverse or
the longitudinal parts respectively of the vacuum polarization
functions that appear in the expression for $\Delta\rho$, and
includes the consideration of non-perturbative effects.

When the two dispersion relations are used to calculate
the effect of the $\ttbar$ threshold on $\Delta\rho$,
they give results which are equal in magnitude but opposite in sign.
This disagreement should not be interpreted as a sign that one of
the dispersion relations is wrong, but as a sign
that neglecting the non--threshold
region when calculating non--perturbative effects to $\Delta\rho$
is a poor approximation.

\bigskip
\centerline{\bf ACKNOWLEDGMENTS}
\bigskip

We would like to thank Profs.
W. A. Bardeen,
E. Braaten,
P. Gambino,
M. C. Gonzalez--Garcia,
F. Halzen,
B. A. Kniehl,
J. L. Rosner,
and
R. A. V\'azquez
for helpful discussions.
This work was supported in part by
the United States Department of Energy under
Grant Number DE-FG02-90ER40560 and
Contract Number DE-AC02-76CH03000.

\end{document}